\documentclass[12pt]{article}
\usepackage{epsfig}
\begin{document}
\title{Absorption of 1$P$-wave heavy charmonium $\chi_{c1}(1P)$ in nuclei}
\author{E. Ya. Paryev\\
{\it Institute for Nuclear Research of the Russian Academy of Sciences}\\
{\it Moscow, Russia}}

\renewcommand{\today}{}
\maketitle

\begin{abstract}
We study the inclusive heavy charmonium $\chi_{c1}(1P)$ photoproduction  from nuclei near the kinematic threshold within the collision model, based on the nuclear spectral function, for incoherent direct photon--nucleon charmonium creation processes. The model accounts for the final $\chi_{c1}(1P)$ absorption in nuclear medium, target nucleon binding and Fermi motion. We calculate the absolute and relative excitation functions on $^{12}$C and $^{184}$W target nuclei at near-threshold photon beam energies of 8.25--16.0 GeV, the absolute momentum differential cross sections and ratios of them for its production off these target nuclei at laboratory polar angles of 0$^{\circ}$--10$^{\circ}$ and for photon
energy of 13 GeV as well as the A-dependences of the transparency ratios for the $\chi_{c1}(1P)$ at photon energy of 13 GeV within the different scenarios for its absorption cross section in nuclei.
We demonstrate that the absolute and relative observables considered reveal distinct sensitivity to these scenarios. Therefore, they might be useful for the determination of this cross section from the comparison of them with the experimental data from the future experiments at the upgraded up to 22 GeV CEBAF facility, which is of crucial importance in understanding of charmonium production and suppression in high-energy heavy--ion collisions in a search for the quark-gluon plasma.
\end{abstract}

\newpage

\section*{1. Introduction}

\hspace{1.5cm} The study of the production and suppression of charmonium states -- the bound states of a heavy charm quark $c$ and its antiquark ${\bar c}$, such as stable under strong decays the 1$S$ vector state $J/\psi$, three 1$P$ states $\chi_{cJ}(1P)$ (with $J=0,1,2$)
\footnote{$^)$In what follows, we will refer to them as $\chi_c$ .}$^)$
and the 2$S$ vector state $\psi(2S)$, whose masses are below the open charm threshold (i.e., $m_{c{\bar c}} < 2m_D$), on a nuclear targets has received considerable experimental and theoretical interest in the last few decades and remains a hot topic in high-energy proton--nucleus and nucleus--nucleus collisions, especially in the hope to observe in these collisions the phase transition at the critical temperature $T_c$ from composite colour-neutral hadrons to a new state of matter -- plasma of deconfined colour-charged quarks and gluons, commonly referred to as the quark-gluon plasma (QGP) [1--10]. Studying this phenomenon and the properties of the formed QGP, using charmonia as probes of them, is important for understanding the very early Universe in period of time a few microseconds after the 'Big Bang' and the structure of the core of neutron stars, which may consist of deconfined quarks. It is the mainstream of the high energy heavy-ion
physics. According to [11--18], the charmonium states produced in a primary nucleon--nucleon collisions, while traversing a hot QGP, should dissociate in it since $c{\bar c}$ pairs, from which they are composed, become unbound due to the effect of color Debye screening of the linear confining interaction between $c$ and ${\bar c}$ quarks. As a result, their yields in the high-energy heavy-ion collisions should be suppressed relative to the scaled proton--proton rates
\footnote{$^)$This determines a great interest in studying experimentally the charmonia production in relativistic proton--proton collisions as a baseline for understanding $pA$ and $AA$ collisions (see, for instance, Refs. [19--21]).}$^)$,
which makes the $c{\bar c}$ bound states relevant probes of the formation of QGP in the first of them. The onset of suppression for different charmonium states is correlated  with the temperature or the energy density of the QGP. With increasing temperature these states "melt" sequently as function of their binding energy: the most loosely bound states $\chi_c$ and $\psi(2S)$ disappear first, the tightly bound ground state $J/\psi$ last. It is expected [17] that the first states "melt" at temperatures around $T_c$, i.e., for energy densities $\sim$ 1 GeV/fm$^3$. In contrast, the $J/\psi$ survives up to $T \ge 2T_c$, which implies to energy densities $\sim$ 25 GeV/fm$^3$ or more. In view of this, we should in principle observe in $AA$ collisions with increasing temperature a sequential step-wise suppression pattern with several drops in the $J/\psi$ survival probability, associated with the dissociation in a hot QGP of the higher excited states $\chi_c$ and $\psi(2S)$ at temperatures just above $T_c$ and the directly produced $J/\psi$ at higher temperatures. The size  of the first drops is determined by the $J/\psi$ feed-down contributions from radiative $\chi_c \to {J/\psi}\gamma$ and hadronic $\psi(2S) \to {J/\psi}X$ decays
\footnote{$^)$It should be noted that the observed $J/\psi$'s feed-down rates from $\chi_c(1P)$ and $\psi(2S)$ decays in
hadron-hadron collisions are approximately 30\% and 10\% and about 60\% of them are produced directly [16, 17, 22, 23].}$^)$.
Therefore, the suppression of $J/\psi$ production in high-energy nucleus--nucleus and proton--nucleus collisions observed at SPS [24, 25], at RHIC [1--5, 8--10, 26, 27] and at LHC [28--30] and the suppression of the production of $\psi(2S)$ mesons relative to that of $J/\psi$ states in large collision systems [9, 31--35] can be considered [1--5, 9--18] as a possible indicator of the formation of QGP at least in these systems and as a sign of their dissociation in it.
Since the $P$-wave charmonia $\chi_{cJ}(1P)$ (with $J=0,1,2$) give essential feed-down contributions to the direct
$J/\psi$ production, the study of their production in nuclear collisions is also very important. Experimentally,
they have not been studied as extensively as the charmonium states $J/\psi$ and $\psi(2S)$.
The measurements of the conventional $\chi_{c1}$ and $\chi_{c2}$ production and their relative amounts have been recently performed at the LHC using proton--proton collisions at center-of-mass energy of 7 TeV by the CMS [36], ATLAS [37] and LHCb [38--40] Collaborations. First evidence for the $\chi_{c0}$ production at a high-energy hadron collider was also presented in Ref. [40]. This state has been previously observed also in $p{\bar p}$ collisions by the E835 Collaboration [41]. Previous measurements of the hadroproduction of the $\chi_{c1}$ and $\chi_{c2}$ mesons have been made using different particle beams, targets and energies (see, for example, Refs. [42--45])
\footnote{$^)$It should be pointed out that a more complete list of previous measurements in this research field is given in Table II from Ref. [45].}$^)$.
The study of the multi-body and isospin-violating decay modes of the $\chi_{c}$ states abundantly produced through the
radiative decays of $\psi(2S)$, collected with the BESIII detector, has been carried out very recently
in Refs. [46, 47]. Interestingly, some theoretical models predict the existence of the first excited state $\chi_{c1}(2P)$ of the $\chi_{c1}(1P)$ (cf. Ref. [48] and references herein), which is considered in the
literature (cf. Ref. [49]) as one of the possible explanations of the famous $X(3872)$ resonance.
It is also worth noting that a sequential step-wise suppression of $J/\psi$, analogous to
that in a hot medium due to the Debye screening of the binding force between $c$ and ${\bar c}$ quarks (the so-called
static suppression [9]), may arise in a dense nuclear matter when with increasing its density the in-medium open charm $D{\bar D}$ mass threshold sequently drops below the in-medium masses of excited charmonium states $\psi(2S)$, $\chi_{c2}(1P)$, $\chi_{c1}(1P)$ and $\chi_{c0}(1P)$, thereby leading to a step-wise suppression of the $J/\psi$ due to the successive "melting" of these states at finite density at the expense of their decays into the $D{\bar D}$ mode [50]. One may hope that this phenomenon could be explored in the upcoming intermediate-energy heavy-ion collision experiments such as the CBM experiment at FAIR (GSI) [51] and the MPD experiment at NICA (Dubna) [52] accelerator facilities. In the QGP, charmonium $J/\psi$ is expected to experience besides the static dissociation described above also the so-called dynamical dissociation [16, 53--56] induced by their inelastic interactions with the QGP constituents -- gluons and quarks, leading to further suppression of its yield in heavy-ion collisions relative to the proton--proton collisions.
On the other hand, another competing mechanisms can also affect the charmonium production process and the overall final rates, including its regeneration via the coalescence of deconfined uncorrelated charm quarks and antiquarks [9,57--60], which becomes more and more important as collision energy increases [9, 16, 17, 60], charmonium dissociation by interactions with hadronic comovers subsequently formed in the collision, i.e., with comoving particles (with light $\pi$, $\rho$, $K$ mesons  in the hadron gas) formed at the late stages of heavy-ion collisions when a QGP expands, cools and hadronizes [61, 62] and with nucleons belonging to the remnant of the colliding nuclei.
These processes need to be accounted for in the consistent description of the charmonium production and suppression in high-energy heavy-ion collisions and in the interpretation of all the existing experimental data.
Thus, it is of great importance for the diagnostics of a QGP formation in these collisions to establish in particular the approximate size of the relatively low-energy
\footnote{$^)$Since the relative motion between the comoving charmonium and nuclear matter is rather slow.}$^)$
$c{\bar c}$ + light hadron/nucleon cross sections [61--63].

Since we have no charmonium beams or targets, the ${J/\psi}N$ "experimental" cross sections are usually inferred indirectly from ${\gamma}p \to {J/\psi}p$ data assuming the vector meson dominance hypothesis [64, 65], or from data on $J/\psi$ nuclear attenuation in ${\gamma}A$ [66] and $pA$ [25, 67, 68] reactions. Since the mid 1970s, the charmonium--nucleon interaction has been intensely debated. However, a full understanding of this interaction is still missing. Thus, the authors of Refs. [64, 65] have used $J/\psi$ photoproduction data on the nucleon combined with a vector meson dominance model find a value of about 3--4 mb for the ${J/\psi}N$ total cross section at c.m.s. energies $\sim$ 10 GeV. The large ${J/\psi}N$ inelastic cross section in the range of 6--8 mb has been estimated in Ref. [63] adopting the effective Lagrangians. The $J/\psi$ absorption cross section in nuclear matter deduced from different experiments clusters around a value of about 3--7 mb. The value of this cross section $\sigma_{{J/\psi}N}=$ ($3.5\pm0.8)$ mb was extracted from the measured $A$-dependence of $J/\psi$ absorption in the SLAC photoproduction experiment [66] at photon energies $\sim$ 20 GeV. The SPS analysis [25], based on $pA$ collisions, gives $\sigma_{{J/\psi}N}=$ ($4.18\pm0.35)$ mb.
An analysis [67, 68] of available data for proton--nucleus collisions at moderate energies ($\sqrt{s_{NN}} \approx 20$ GeV) leads to a value of about 6--7 mb of the $J/\psi$ absorption cross section on a nucleon, which is larger than that deduced in the SLAC experiment [66] by a factor $\approx$ 2.
On the other hand, information on the $\psi(2S)$--nucleon and $\chi_{cJ}(1P)$--nucleon interactions is scarce in the literature and, in particular, the total cross sections for these interactions are poorly known.
Thus, the SPS analysis [25] of the proton--nucleus $\psi(2S)$ production data gives the value of (7.3$\pm$1.6) mb
for the $\psi(2S)$--nucleon absorption cross section, which is greater than that for the $J/\psi$ extracted
in the SLAC experiment [66] by a factor of about two. Theoretically,
the assumption that the charmonium--nucleon total cross section scales with the charmonium spatial size squared
[64, 69--72] leads to the conclusion that $\psi(2S)N$ total cross section is  a factor of about 4 larger than the ${J/\psi}N$ one due to the larger size of the $\psi(2S)$ compared to that of the $J/\psi$ [64, 70] and it may reach the values about of 20 mb at the $\psi(2S)N$ center-of-mass energy of 10 GeV [71]. Under the charmonium size we mean the root-mean-square size of the separation between charm quark and antiquark in a charmonium -- the r.m.s. radius of charmonium.
Quark potential models [73--77] find the r.m.s. $c{\bar c}$ separations of about 0.4 fm for the $J/\psi$, 0.6 fm for the $\chi_c$ states and 0.8 fm for the $\psi(2S)$. In line with the above "scaling" assumption, we have yet that the ${\chi_c}N$ total cross section is approximately 2.25 times  larger than corresponding ${J/\psi}N$ cross section (cf. Ref. [70]). This leads to its value of about 8 mb for the $J/\psi$--nucleon cross section $\sigma_{{J/\psi}N}=$ 3.5 mb [66]. Using the QCD factorization theorem and the non-relativistic charmonium model with a Cornell confining potential, the authors of Ref. [71] have calculated the total interaction cross sections $\sigma_0^{\chi}$ and $\sigma_1^{\chi}$ of a $c{\bar c}$ $P$-wave state with orbital magnetic quantum numbers $m=0$ and $m=\pm1$ at $(c{\bar c})N$ c.m. collision energy of 10 GeV. They are: $\sigma_0^{\chi}=6.8$ mb and $\sigma_1^{\chi}=15.9$ mb. However, the predicted in [72] cross sections at this energy for $\chi_c(m=0)$ and $\chi_c(m=1)$ are about twice as less as the above ones. The total cross section of a physical $\chi_{cJ}(\nu)$ state ($J=0,1,2$) with helicity $\nu$--nucleon can be expressed in terms of the two cross sections $\sigma_0^{\chi}$ and $\sigma_1^{\chi}$ (see relations (41) from Ref. [72]). Adopting these relations, one can get that the averaged over helicities $\chi_{cJ}(1P)N$ total cross sections are equal for all three states $\chi_{c0,1,2}$ [72]. Throughout the paper we refer to these three degenerate cross sections as $\sigma_{{\chi_c}N}$.
This quantity is expressed through the cross sections $\sigma_0^{\chi}$ and $\sigma_1^{\chi}$ as follows:
$\sigma_{{\chi_c}N}=\frac{1}{3}(\sigma_0^{\chi}+2\sigma_1^{\chi})$ and has the values of about 13 mb and 6.5 mb for the above-mentioned predictions for these cross sections from Refs. [71] and [72], respectively, at c.m. energy of 10 GeV
relevant for our work. The numerical results for the low-energy dissociation cross sections of conventional ground-state, orbitally- and radially-excited charmonia in collisions with light mesons ($\pi$, $\rho$, $K$) were presented in Refs. [78, 79]. These cross sections are important for the study of heavy-ion collisions. In particular, the 2--4 mb scale cross sections were found here for the low-energy ${\chi_{cJ}}\pi$ interactions. Since
the r.m.s. radius of the full-sized $\chi_c$ is less than that of the $\psi(2S)$ and is larger than that of the $J/\psi$, we can expect in line with Refs. [64, 69--72] that for the same c.m. kinetic energy the ${\chi_c}N$ unpolarized total cross section $\sigma_{{\chi_c}N}$ may be smaller than the ${\psi(2S)}N$ total cross section $\sigma_{{\psi(2S)}N}$ and may be greater than the ${J/\psi}N$ total cross section $\sigma_{{J/\psi}N}$, i.e. the following ordering among these free-space cross sections should be
\begin{center}
$\sigma_{{J/\psi}N} < \sigma_{{\chi_c}N} < \sigma_{{\psi(2S)}N}$,
\end{center}
which is inversely correlated also with the $J/\psi$, $\chi_c$, $\psi(2S)$ binding energies $E_{J/\psi}$, $E_{\chi_c}$, $E_{\psi(2S)}$ hierarchy
\begin{center}
$E_{J/\psi} > E_{\chi_c} > E_{\psi(2S)}$
\footnote{$^)$The binding energy of the $\psi(2S)$ ($\approx$ 0.05 GeV) is significantly smaller than that of the
$\chi_c$ ($\approx$ 0.20 GeV) or $J/\psi$ ($\approx$ 0.64 GeV) [16, 17]. They are defined as the differences between the open charm threshold and charmonia masses.}$^)$.
\end{center}
In view of the aforesaid, we will adopt in our calculations the following four representative options for the in-medium
(effective) full-sized ${\chi_c}$--nucleon absorption cross section $\sigma_{{\chi_c}N}$, namely: 3.5, 7, 14 and 20 mb, covering the bulk of the free-space low-energy theoretical information presently available in this field
\footnote{$^)$On the other hand, since the medium effects are expected to be minimal for the relatively high $\chi_{c1}(1P)$ momenta relevant for our study (see below), we can assign the values of $\sigma_{{\chi_c}N}$ deduced from an analysis of $\chi_{c1}(1P)$ photoproduction off nuclei within the present model to those characterizing a real cross section for heavy charmonium absorption by nucleons in the free space [80].}$^)$.

To discriminate between these options and to gain further insights into the low-energy $\chi_{c0,1,2}(1P)$ meson--nucleon interaction in cold nuclear matter, it is extremely important and timely, according to the above, to study in particular the photoproduction of unpolarized $\chi_{c1}(1P)$ mesons
\footnote{$^)$Having the largest branching ratio for the radiative $\chi_{c1}(1P) \to {J/\psi}{\gamma}$ decays (34.3\% as opposed to 1.4\% for $\chi_{c0}(1P)$ and 19.5\% for $\chi_{c2}(1P)$ [81]).}$^)$
on protons and nuclei at energies close to the threshold for their production off a free nucleon. This has the advantage compared to the high-energy hadronic collisions that the interpretation of data from photoproduction experiments on nuclei is clearer due to a negligible strength of initial-state photon interaction and since their production proceeds through a few elementary channels in a cleaner environment - in static cold nuclear medium whose density is sufficiently well known. Moreover, in elementary low-energy photon-induced reactions, contrary to the high-energy ones, the $\chi_{c1}(1P)$ mesons are produced with relatively low momenta in the target nucleus rest-frame at which the coherence and formation length effects play inessential role and the final full-sized $\chi_{c1}(1P)$ interacts with nuclear matter, but not the "premeson" $c{\bar c}$ (see below). Recently, the GlueX Collaboration at the JLab has reported the observation of 56.5$\pm$8.2 $\chi_{c1}(1P)$ and 12.7$\pm$4.5 $\chi_{c2}(1P)$ production events in the exclusive ${\gamma}p \to {\chi_{c1}}p$ and ${\gamma}p \to {\chi_{c2}}p$ reactions at threshold energies close to the end point of the photon spectrum of 11.4 GeV [82]. Here, the charmonium states $\chi_{c1}(1P)$ and $\chi_{c2}(1P)$ were detected by their radiative decays to $J/\psi$: $\chi_{c1}(1P) \to {J/\psi}{\gamma} \to e^+e^-{\gamma}$ and $\chi_{c2}(1P) \to {J/\psi}{\gamma} \to e^+e^-{\gamma}$.
A comprehensive studies of such higher-mass charmonia (and $\psi(2S)$ particles) are planned [82] to be performed at the upgraded up to 22 GeV CEBAF facility [83--85] with a photon beam with higher intensity in the energy region near the $J/\psi$ threshold and in the high-energy region above 12 GeV
\footnote{$^)$It is worth mentioning that the near-threshold photoproduction of $J/\psi$ on the proton has been recently studied by the GlueX [86, 87] and $J/\psi$--007 [88] experiments at the JLab. Moreover, the first measurement of near and subthreshold $J/\psi$ photoproduction off deuterium, helium and carbon target nuclei, using the GlueX spectrometer, has been reported in the recent publication [89]. However, in view of the low statistics achieved in this measurement, a new higher-statistics experiment on a liquid $^4$He target is planned to be performed at the JLab by the GlueX Collaboration [90].}$^)$.
Theoretically, the study of the photoproduction of unpolarized $\chi_{c1}(1P)$ mesons off a proton target has been carried out in Refs. [91, 92]. Their exclusive and inclusive production cross sections from photon and vector meson exchanges have been predicted at center-of-mass energies $W$ of the photon-proton system $W \le 7$ GeV relevant for the present study. It was found that at these energies the dominant contribution comes from the $\omega$ exchanges, while the contribution of photon exchange is negligible. The exclusive photoproduction of $P$-wave spin-triplet heavy charmonia $\chi_{cJ}$ (with $J=0,1,2$) via one-photon exchange (a Primakoff process) and through one-photon plus three gluon (Odderon) exchanges from the proton in high-energy ${\gamma^*}p$ collisions has been investigated in Refs. [93] and [94], respectively. Should be said, the aforementioned photon exchange serves as a background to the Odderon exchange, which, as is expected [94], dominates over the Primakoff process at high momentum transfer ($|t| \gg 1$ GeV$^2$) in the differential cross sections. This fact may give a chance to find a strong evidence for the Odderon exchange model from the future measurements of the exclusive electroproduction of $P$-wave charmonia $\chi_{cJ}(1P)$, with $J=0,1,2$, at the under-construction high-luminosity electron-ion colliders EIC [95, 96] and EicC [97, 98] in the United States and China.
At present there are no measurements of the low-energy $\chi_{c1}(1P)$ production and absorption on nuclei in near-threshold photon-induced reactions.

To inspire them, in this study we present the detailed predictions for the absolute and relative excitation functions for production of unpolarized $\chi_{c1}(1P)$ mesons off $^{12}$C and $^{184}$W target nuclei, for the absolute momentum distributions and ratios of them for their production off these target nuclei as well as for the A-dependences of the transparency ratios for the $\chi_{c1}(1P)$ mesons from ${\gamma}A$ reactions at threshold energies obtained within the first collision model assuming the above realistic scenarios for their in-medium absorption cross section. Our predictions can be tested by future measurements at the upgraded up to 22 GeV CEBAF facility with the aim of discriminating between these scenarios. This would clearly be of great importance both for the relevance to QGP searches and as a valuable test of the theoretical predictions in the field of charmonia scattering.

\section*{2. Direct $\chi_{c1}(1P)$ photoproduction mechanism: cross sections and their ratios}

\hspace{1.5cm} Direct $\chi_{c1}(1P)$ photoproduction on nuclear targets at the near-threshold laboratory incident photon energies $E_{\gamma} \le 16$ GeV of interest
\footnote{$^)$Which corresponds to the center-of-mass energies $W$ of the photon-proton system $W \le 5.56$ GeV,
or to the relatively "low" excess energies $\epsilon$ above the $\chi_{c1}(1P)p$ production threshold
$W_{\rm th}=\sqrt{s_{\rm th}}=m_{\chi_{c1}}+m_{p}=$ 4.44895 GeV ($m_{\chi_{c1}}$ and $m_{p}$ are the $\chi_{c1}(1P)$ meson and proton free space masses, respectively ) $0 \le \epsilon \le 1.11$ GeV and where the $\chi_{c1}(1P)$ mesons can be observed in the ${\gamma}p$ and ${\gamma}A$ reactions at the upgraded up to 22 GeV
CEBAF facility at the JLab [83--85]. We remind that the measured mass, full width and quantum numbers of the $\chi_{c1}(1P)$, respectively, are $m_{\chi_{c1}}=(3510.67\pm0.05)$ MeV, $\Gamma_{\chi_{c1}}=(0.84\pm0.04)$ MeV and $J^{PC}=1^{++}$ [81].}$^)$
may proceed via the following elementary processes with the lowest free production threshold ($\approx$ 10.08 GeV)
[91, 92]:
\begin{equation}
{\gamma}+p \to \chi_{c1}(1P)+p,
\end{equation}
\begin{equation}
{\gamma}+n \to \chi_{c1}(1P)+n.
\end{equation}
In passing, we note that according to Ref. [92], we can ignore the processes ${\gamma}N \to \chi_{c1}(1P)N{\pi}$ with one pion in the final states at these energies. For the $\chi_{c1}(1P)$ charmonium (denoted below as the $\chi_{c1}$) there is also a feed-down channel ${\gamma}N \to \psi(2S)N \to \chi_{c1}(1P){\gamma}N$ from exclusive production of $\psi(2S)$ on nucleon, taking place above the threshold energy of 10.93 GeV, with subsequent radiative decay $\psi(2S) \to {\chi_{c1}(1P)}\gamma$. Although the $\psi(2S)$ and $\chi_{c1}(1P)$ total cross sections are of similar magnitude at above threshold photon energies (see Fig. 1 below), the small $\psi(2S)$ branching ratio, $\approx$ 8.8\%, would result in its negligible contribution to $\chi_{c1}(1P)$ production. Moreover, in the energy domain of our interest we can neglect the following two-step $\chi_{c1}(1P)$ production processes with $\psi(3770)$ and $X(3872)$ mesons in an intermediate states and their subsequent decays into the $\chi_{c1}(1P)$: ${\gamma}N \to \psi(3770)N$, ${\gamma}N \to X(3872)N$; $\psi(3770) \to \chi_{c1}(1P){\gamma}$, $X(3872) \to \chi_{c1}(1P){\pi^0}$, $X(3872) \to \chi_{c1}(1P){\pi}{\pi}$ due to larger $\psi(3770)$ and $X(3872)$ production thresholds in ${\gamma}N$ collisios--11.36 and 11.86 GeV, respectively, and owing to very small branching ratios of these decays -- 2.9$\cdot$10$^{-3}$ [81], 3.8\% [99] and of the order of 10$^{-3}$ [99], correspondingly.
The $\chi_{c1}(1P)$ mesons and nucleons, produced in these processes, are sufficiently energetic.
Thus, for example, the kinematically allowed $\chi_{c1}(1P)$ meson and final
proton laboratory momenta in the direct process (1), proceeding on the free target proton at rest, vary within the momentum ranges of 6.654--12.232 GeV/c and 0.768--6.346 GeV/c, respectively, at incident photon energy of
$E_{\gamma}=13$ GeV. Since the neutron mass is approximately equal to the proton mass,
the kinematical characteristics of final particles, produced in the reaction (2), are similar to those of
final particles in the process (1). Since the medium effects are expected to be reduced for high momenta,
we will ignore the medium modifications of the outgoing sufficiently energetic $\chi_{c1}(1P)$ mesons and
nucleons in the case when the reactions (1), (2) proceed on a nucleons embedded in a nuclear target
\footnote{$^)$It should be pointed out that the authors of Refs. [100, 101] and [102] adopting, respectively, the QCD sum rule calculations and the Quark-Meson-Coupling model have found nearly degenerate mass shifts of about -60 MeV
for $\chi_{cJ}(1P)$ ($J=0,1,2$) quarkonia at normal nuclear matter density $\rho_0$ and low energies, which amounts approximately to 1.7\% of their free-space nominal masses.}$^)$.

Disregarding the absorption of incident photons in the energy range of interest [103] and assuming instantaneous production of the full-sized $\chi_{c1}(1P)$ meson on a single bound nucleon
\footnote{$^)$This assumption is valid when the so-called coherence length $l_c$--the distance that the virtual
small-sized $c{\bar c}$ fluctuation of the incoming photon travels in the lab frame before scattering on this nucleon--is much less than the average inter-nucleon distance and there is no need to add coherently the production amplitudes on the two nucleons separated by the distance less than $l_c$ [71, 72, 104, 105]. Accounting for that the coherence length
$l_c=2E_{\gamma}/m^2_{\chi_{c1}}$ [72, 104--106], we get that at the photon energy $E_{\gamma}=13$ GeV, accessible at the JLab upgraded up to 22 GeV, it is about of 0.4 fm. This value is essentially less than the average inter-nucleon distance of about 2 fm. Therefore, in this case the final compact $c{\bar c}$ pair is created substantially right at the location of the target nucleon that it scatters from. After this scattering it evolves over some
time (the formation time $t_f$) or over some distance (the formation length $l_f$) into a final full-sized
$\chi_{c1}$ meson [71, 72, 104, 105]. In the case when the formation length is much less than the nuclear radius, we can assume that such full-sized meson propagates from its creation point through the nucleus as ordinary hadron and its absorption during this propagation is described by the $\chi_{c1}$--nucleon absorption cross section $\sigma_{\chi_{c}N}$ defined above. For the $\chi_{c1}$ we apply the estimate of Ref. [71]: $l_f=3{\rm fm\frac{p_{\chi_{c1}}}{30{\rm GeV}}}$, where $p_{\chi_{c1}}$ is the momentum of the $\chi_{c1}$ in the rest frame of the target. With $p_{\chi_{c1}} \approx 10$ GeV/c, the average momentum of a $\chi_{c1}$ produced in the process (1)
on the free target proton at rest at the photon energy of 13 GeV (see above), this yields $l_f \approx 1$ fm.
It is close to the nucleon size than to the nucleus size (thus, the radii of $^{12}$C and $^{184}$W target nuclei are approximately 3 and 7.4 fm, respectively). This favors the determination of the real free-space absorption cross section of a charmed $\chi_{c1}$ meson on a nucleon, in contrast with high energies which instead allow for access only to the nuclear interaction of a small-sized compact embryonic $c{\bar c}$ pair.}$^)$,
we can represent the total cross section for the production of unpolarized $\chi_{c1}(1P)$ mesons
on nuclei from the direct photon--induced reaction channels (1), (2) as follows [107]:
\begin{equation}
\sigma_{{\gamma}A\to \chi_{c1}X}^{({\rm dir})}(E_{\gamma})=I_{V}[A,\sigma_{\chi_{c}N}]
\left<\sigma_{{\gamma}p \to \chi_{c1}p}(E_{\gamma})\right>_A,
\end{equation}
where the effective number of target nucleons participating in the direct processes (1), (2), $I_{V}[A,\sigma_{\chi_{c}N}]$, and "in-medium" total cross section for the production of $\chi_{c1}(1P)$ mesons
in reaction (1) $\sigma_{{\gamma}p \to \chi_{c1}p}(\sqrt{s^*})$ at the in-medium ${\gamma}p$ center-of-mass energy $\sqrt{s^*}$, averaged over target nucleon binding and Fermi motion, $\left<\sigma_{{\gamma}p \to \chi_{c1}p}(E_{\gamma})\right>_A$, are defined by Eqs. (4), (5) and (6) from Ref. [107], respectively,
in which one needs to make the substitution: $X(3872) \to \chi_{c1}$
\footnote{$^)$In Eq. (3) we assume that the $\chi_{c1}(1P)$ meson production
cross sections in ${\gamma}p$ and ${\gamma}n$ interactions are the same
and neglect the difference between proton ($m_p$) and neutron ($m_n$) masses.}$^)$.

  As before in Ref. [107], we suggest here that the "in-medium" cross section
$\sigma_{{\gamma}p \to \chi_{c1}p}({\sqrt{s^*}})$ for $\chi_{c1}(1P)$ production in process (1)
is equivalent to the vacuum cross section $\sigma_{{\gamma}p \to \chi_{c1}p}({\sqrt{s}})$,
in which the free space center-of-mass energy squared $s$ for given photon laboratory energy $E_{\gamma}$
and momentum ${\bf p}_{\gamma}$, presented by the formula
\begin{equation}
s=s(E_{\gamma})=W^2=(E_{\gamma}+m_p)^2-{\bf p}_{\gamma}^2=m_p^2+2m_pE_{\gamma},
\end{equation}
is replaced by the in-medium expression
\begin{equation}
  s^*=(E_{\gamma}+E_t)^2-({\bf p}_{\gamma}+{\bf p}_t)^2,
\end{equation}
\begin{equation}
   E_t=M_A-\sqrt{(-{\bf p}_t)^2+(M_{A}-m_{p}+E)^{2}}.
\end{equation}
Here, $E_t$, ${\bf p}_{t}$ and $E$ are the total energy, momentum and binding energy of the struck target
proton involved in the collision process (1).
For the free total cross section $\sigma_{{\gamma}p \to {\chi_{c1}}p}(\sqrt{s})$ of the reaction (1) no data are available presently at the considered photon energies $E_{\gamma} \le $ 16 GeV relevant for the upgraded to the energy of 22 GeV CEBAF facility. Therefore, we have to rely on some theoretical predictions and estimates for it, existing in the literature at these energies. For this cross section we have used the following parametrization of the results of its calculations here within the vector meson exchange model [91, 92]
\begin{equation}
\sigma_{{\gamma}p \to {\chi_{c1}}p}(\sqrt{s})=0.427\left(1-\frac{s_{\rm th}}{s}\right)^{1.047}\left(\frac{s}{s_{\rm th}}\right)^{0.453}~[\rm nb],
\end{equation}
where
\begin{equation}
s_{\rm th}=(m_{\chi_{c1}}+m_{p})^2=(4.44895~{\rm GeV})^2.
\end{equation}
The results of calculations by Eq. (7) of the total cross section of the reaction
${\gamma}p \to {\chi_{c1}}p$ at laboratory photon energies $E_{\gamma} \le 16$ GeV, are shown in Fig. 1 (solid curve).
\begin{figure}[htb]
\begin{center}
\includegraphics[width=15.0cm]{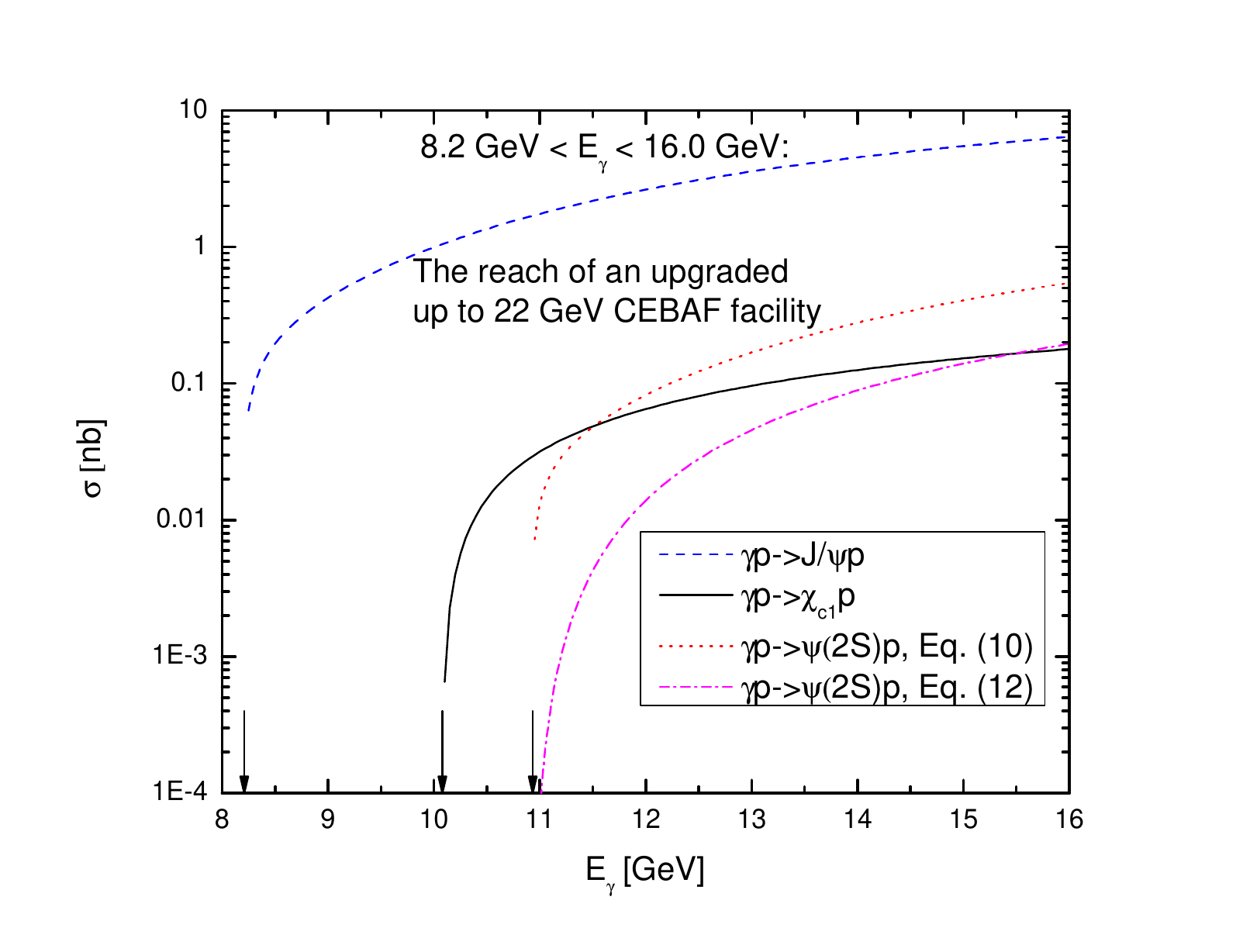}
\vspace*{-2mm} \caption{(Color online.) Total cross sections for the reactions ${\gamma}p \to {\chi_{c1}(1P)}p$,
${\gamma}p \to {J/\psi}p$ and ${\gamma}p \to {\psi(2S)}p$ as functions of the laboratory energy $E_{\gamma}$ of the photon--proton collisions in the kinematic range accessible at the upgraded up to 22 GeV JLab facility [83--85].
Solid, dashed, dotted and dashed-dotted curves represent calculations performed
using Eqs. (7), (9), (10) and (12), respectively. The arrows indicates the threshold energies of 8.21, 10.08 and 10.93 GeV for $J/\psi$, $\chi_{c1}(1P)$ and $\psi(2S)$ photoproduction on a free target proton at rest.}
\label{void}
\end{center}
\end{figure}
For comparison with this cross section, we also show in Fig. 1 the results of calculations
of the free total cross section $\sigma_{{\gamma}p \to {J/\psi}p}({\sqrt{{s}}})$ of the
${\gamma}p \to {J/\psi}p$ reaction in this domain (dashed curve) performed using the following parametrization [108] of
the available experimental data [86], based on the predictions of the two gluon and three gluon exchange model
[109] near threshold:
\begin{equation}
\sigma_{{\gamma}p \to {J/\psi}p}({\sqrt{{s}}})= \sigma_{2g}({\sqrt{{s}}})+\sigma_{3g}({\sqrt{{s}}}),
\end{equation}
where 2$g$ and 3$g$ exchanges cross sections $\sigma_{2g}({\sqrt{{s}}})$ and
$\sigma_{3g}({\sqrt{{s}}})$ are given in Ref. [108] by formulas (7) and (8), respectively.
In addition, in this figure we also depict the predictions for the total cross section of the reaction
${\gamma}p \to {\psi(2S)}p$ from the recent parametrization [110]
\begin{equation}
 \sigma_{{\gamma}p \to {\psi(2S)}p}(\sqrt{s})=0.166\sigma_{{\gamma}p \to {J/\psi}p}(\sqrt{{\tilde s}}),
\end{equation}
in which the total cross section $\sigma_{{\gamma}p \to {J/\psi}p}(\sqrt{{\tilde s}})$ is calculated in line
with Eq. (9) at the c.m. energy $\sqrt{{\tilde s}}$ defined as
\begin{equation}
\sqrt{{\tilde s}}=\sqrt{s}-m_{\psi(2S)}+m_{J/\psi}
\end{equation}
(dotted curve) and from the current parametrization [111]
\begin{equation}
 \sigma_{{\gamma}p \to {\psi(2S)}p}(\sqrt{s})=0.674\left(1-\frac{(m_{\psi(2S)}+m_p)^2}{s}\right)^2\cdot(\sqrt{s})^{0.65}~[\rm nb],
\end{equation}
adopted in the STARlight Monte Carlo simulation program package to simulate the $\psi(2S)$ production in ultra-peripheral collisions of relativistic ions (dashed-dotted curve). Here, $m_{\psi(2S)}$ and $m_{J/\psi}$ are the $\psi(2S)$ and $J/\psi$ free space masses, respectively. We remind that the photoproduction of these three charmonium states off the
nuclear targets is planned to be studied with essentially higher statistics by the GlueX Collaboration with the proposed 22 GeV upgrade of the JLab [82, 90].
\begin{figure}[htb]
\begin{center}
\includegraphics[width=15.0cm]{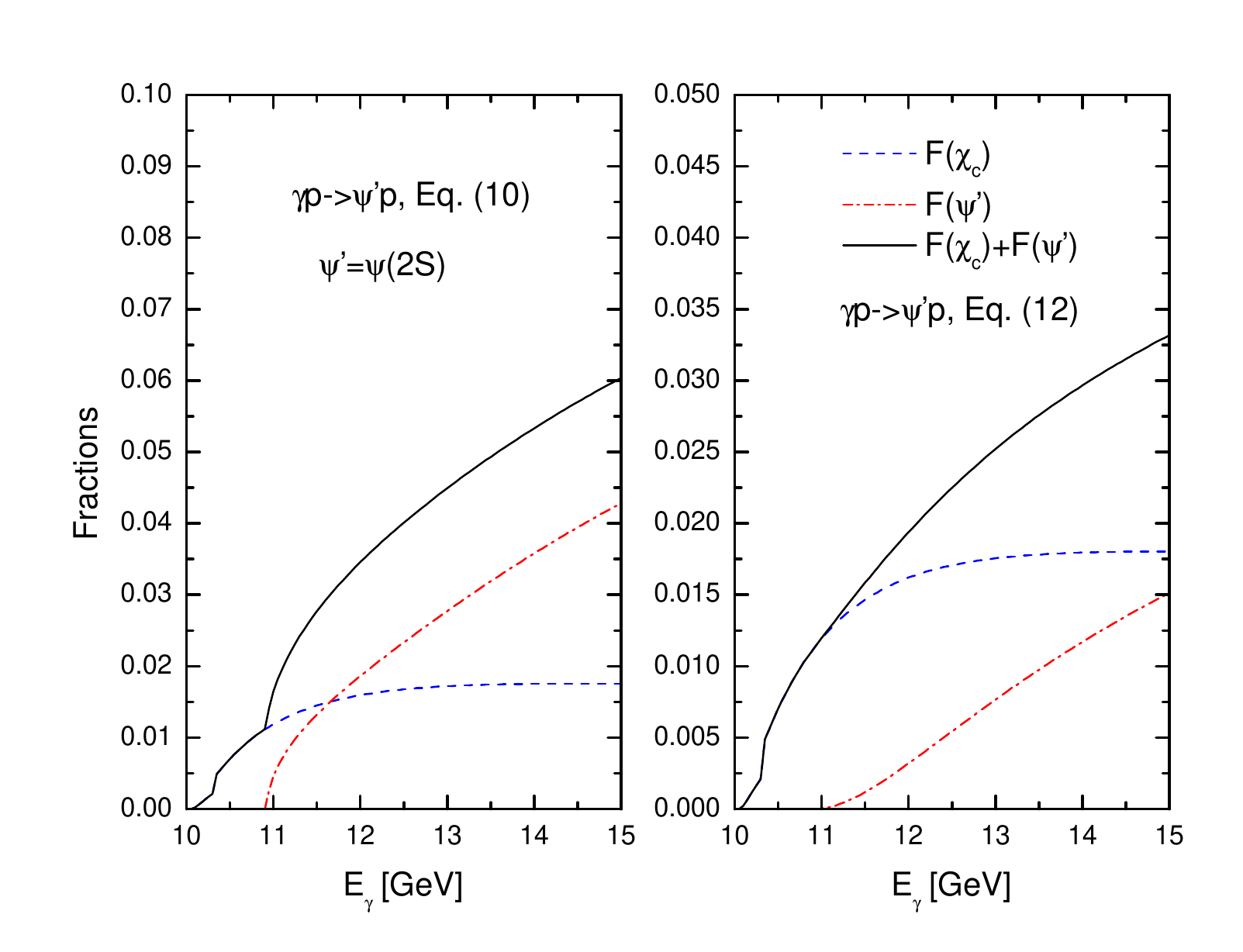}
\vspace*{-2mm} \caption{(Color online.) The fractions of $J/\psi$ mesons arising from radiative decays of the two $\chi_{c1}(1P)$, $\chi_{c2}(1P)$ states ($F(\chi_c)$) and hadronic decays of the $\psi(2S)$ mesons ($F(\psi'))$
produced, respectively, in the direct reactions ${\gamma}p \to {\chi_{c1,c2}(1P)}p$ and ${\gamma}p \to {\psi(2S)}p$ with respect to the total $J/\psi$ yield originating from these decays and from the process
${\gamma}p \to {J/\psi}p$ as well as their sum, $F(\chi_c)+F(\psi')$, as functions of the incident photon energy $E_{\gamma}$. The left and right panels represent calculations performed using, correspondingly, Eqs. (10) and (12) for the total cross section of the ${\gamma}p \to {\psi(2S)}p$ reaction.}
\label{void}
\end{center}
\end{figure}
This figure shows that in the regime of $E_{\gamma} > 11$ GeV the $\chi_{c1}$ and both $\psi(2S)$ cross sections are of similar magnitude, they are of the order of 0.1--0.2 nb for photon energies of 13--15 GeV and are about one order of  magnitude smaller than the $J/\psi$ production cross section at these energies. This hints on an important for the future near-threshold $J/\psi$ photoproduction experiment at the JLab fact: in it the $J/\psi$'s are mostly produced directly but not practically originate from feed-down processes involving $\chi_{c1,c2}(1P)$ and $\psi(2S)$ states.

Indeed, as follows from Fig. 2, where the fractions of $J/\psi$ mesons, $F(\chi_c)$ and $F(\psi')$ coming, respectively, from feed-down decays $\chi_{c1,c2}(1P) \to {J/\psi}{\gamma}$ and $\psi(2S) \to {J/\psi}X$ of the three states $\chi_{c1,c2}(1P)$ and $\psi(2S)$ produced in the elementary reactions ${\gamma}p \to {\chi_{c1,c2}}p$ and ${\gamma}p \to {\psi(2S)}p$ with respect to the total $J/\psi$ yield originating from their direct production in the ${\gamma}p \to {J/\psi}p$ reaction and from the above feed-down processes are shown as functions of photon energy, the $J/\psi$ total feed-down contributions from the $\chi_{c1,c2}$ and $\psi(2S)$ decays, $F(\chi_c)+F(\psi')$, do not exceed 6\% and 3.5\%
\footnote{$^)$Compare with the $(\chi_c+\psi(2S))$-to-$J/\psi$ feed-down fraction of about 40\% measured in high-energy hadronic collisions (see above).}$^)$
at the considered photon energies when the fits (10) and (12) for the $\psi(2S)$ production cross section in ${\gamma}p$ collisions are used in the calculations. The fractions $F(\chi_c)$ and $F(\psi')$ of indirectly produced $J/\psi$'s, presented in Fig. 2, were calculated in line with the formulas (cf. Refs. [23, 45])
\begin{equation}
F(\chi_c)=\frac{{\sum_{J=1}^{J=2}}\sigma_{{\gamma}p \to {\chi_{cJ}}p}(\sqrt{s})Br[\chi_{cJ} \to {J/\psi}\gamma]}{\sigma_{J/\psi}^{\rm tot}},
\end{equation}
\begin{equation}
F(\psi')=\frac{\sigma_{{\gamma}p \to \psi(2S)p}(\sqrt{s})Br[\psi(2S) \to {J/\psi}X]}{\sigma_{J/\psi}^{\rm tot}},
\end{equation}
where
\begin{equation}
\sigma_{J/\psi}^{\rm tot}=
\sigma_{{\gamma}p \to {J/\psi}p}(\sqrt{s})+\sigma_{{\gamma}p \to {\psi(2S)}p}(\sqrt{s})
Br[\psi(2S) \to {J/\psi}X]+\sum_{J=1}^{J=2}\sigma_{{\gamma}p \to {\chi_{cJ}}p}(\sqrt{s})Br[\chi_{cJ} \to{J/\psi}\gamma]
\end{equation}
and the total cross sections $\sigma_{{\gamma}p \to {\chi_{c1}}p}(\sqrt{s})$, $\sigma_{{\gamma}p \to {J/\psi}p}(\sqrt{s})$, $\sigma_{{\gamma}p \to \psi(2S)p}(\sqrt{s})$ are given in Fig. 1, and the branching ratios
$Br[\chi_{c1} \to {J/\psi}\gamma]$, $Br[\chi_{c2} \to {J/\psi}\gamma]$, $Br[\psi(2S) \to {J/\psi}X]$, respectively, are
34.3\%, 19.5\%, 61.5\% [81]. The cross section $\sigma_{{\gamma}p \to {\chi_{c2}}p}(\sqrt{s})$ was evaluated assuming
the consistency of the cross section ratio $\sigma_{{\gamma}p \to {\chi_{c2}}p}(\sqrt{s})/\sigma_{{\gamma}p \to {\chi_{c1}}p}(\sqrt{s})$ with the simple spin-state counting expectation of 5/3 [112, 113].

The local proton and neutron densities, adopted in the calculations of the quantity $I_{V}[A,\sigma_{{\chi_c}N}]$, entering into Eq. (3), for the target nuclei $^{12}_{6}$C, $^{27}_{13}$Al, $^{40}_{20}$Ca, $^{63}_{29}$Cu, $^{93}_{41}$Nb, $^{112}_{50}$Sn, $^{184}_{74}$W, $^{208}_{82}$Pb and $^{238}_{92}$U considered in the present work
are given in Ref. [107]. For medium-weight $^{93}_{41}$Nb, $^{112}_{50}$Sn and heavy $^{184}_{74}$W, $^{208}_{82}$Pb, $^{238}_{92}$U target nuclei we use the neutron density $\rho_n(r)$ in the 'skin' form.

The estimate of the rate of the $\chi_{c1}$ photoproduction on nuclei necessitates also the specifying of the input $\chi_c$--nucleon absorption cross section $\sigma_{{\chi_c}N}$, which determines the quantity $I_{V}[A,\sigma_{{\chi_c}N}]$ (cf. Eqs. (4), (5) from Ref. [107]). In view of the above, in our present study we will adopt the following four main options for this cross section: 3.5, 7, 14 and 20 mb and (sometimes, see Fig. 10 below) several additional ones: 0, 10.5, 25, 30, 35, 40, 45 and 50 mb to extend the range of applicability of our model.

As a measure for the $\chi_{c}$ absorption cross section $\sigma_{\chi_{c}N}$ in nuclei we will use two additional
integral observables.  The first one is the so-called $\chi_{c1}$ transparency ratio defined as (see Ref. [107] and references herein):
\begin{equation}
S_A=\frac{\sigma_{{\gamma}A \to \chi_{c1}X}^{({\rm dir})}(E_{\gamma})}{A~\sigma_{{\gamma}p \to \chi_{c1}p}(\sqrt{s(E_{\gamma})})},
\end{equation}
{\it i.e.} the ratio between the inclusive nuclear $\chi_{c1}(1P)$ photoproduction cross section (3) and by $A$ times the same quantity on a free proton. The second one is the $\chi_{c1}$ transparency ratio $S_A$ normalized to a light nucleus
like $^{12}$C [114, 115]:
\begin{equation}
T_A=\frac{S_A}{S_{\rm C}}=\frac{12~\sigma_{{\gamma}A \to \chi_{c1}X}^{({\rm dir})}(E_{\gamma})}{A~\sigma_{{\gamma}{\rm C} \to \chi_{c1}X}^{({\rm dir})}(E_{\gamma})}.
\end{equation}
The quantities (16) and (17) are the cross-section ratios. They  are sensitive to the $\chi_c$--nucleon absorption cross section. On the other  hand, they are less sensitive than cross sections themselves to the theoretical uncertainties associated mainly with the experimentally unknown total cross sections of the elementary reactions (1), (2).
\begin{figure}[!h]
\begin{center}
\includegraphics[width=15.0cm]{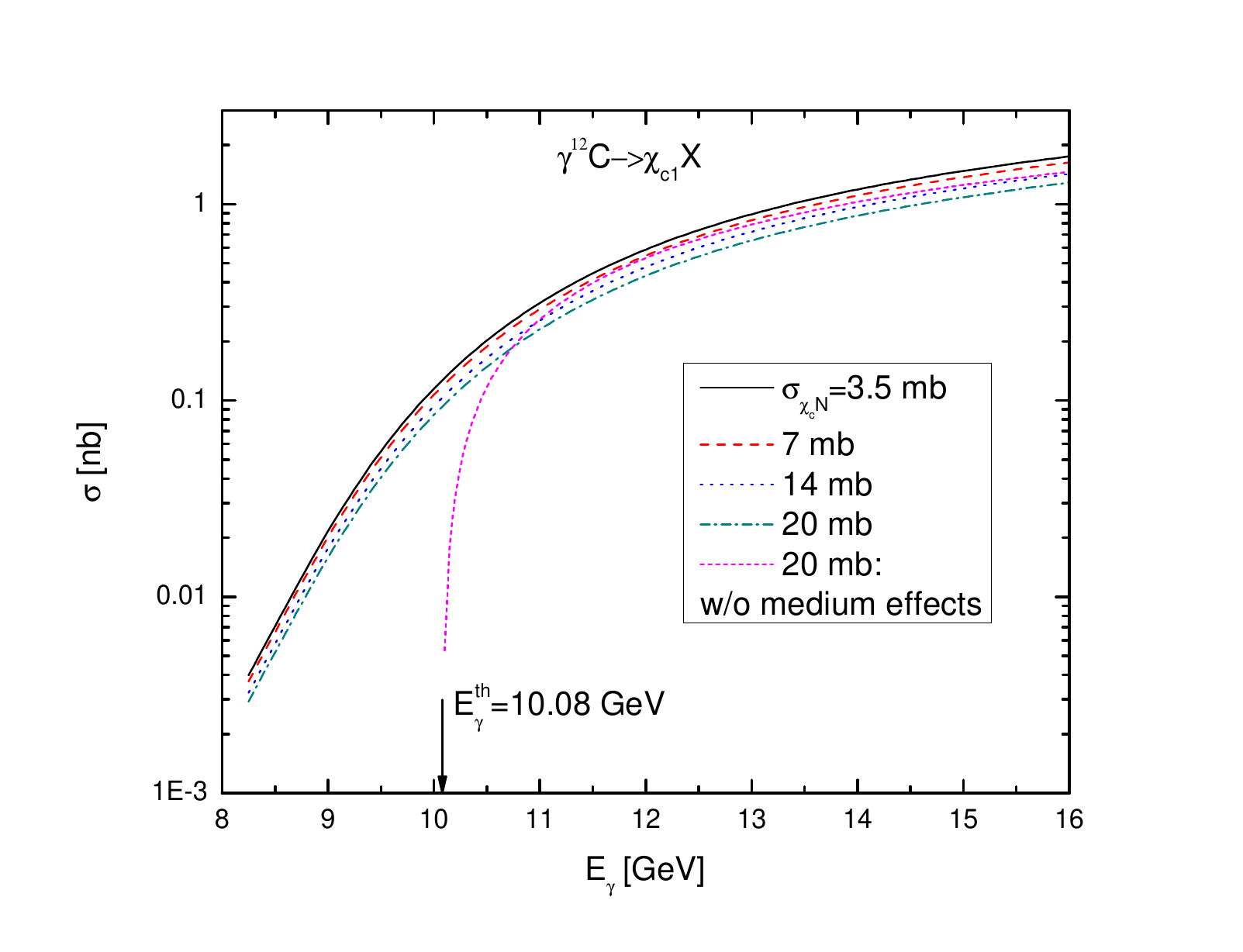}
\vspace*{-2mm} \caption{(Color online.) Excitation function for production of $\chi_{c1}(1P)$
mesons off $^{12}$C from the direct reactions (1), (2) proceeding on an off-shell target nucleons
and on a free ones being at rest. The curves are calculations for $\sigma_{\chi_{c}N}=$ 3.5, 7, 14 and 20 mb.
The arrow indicates the threshold energy for the $\chi_{c1}(1P)$ photoproduction on a free nucleon.}
\label{void}
\end{center}
\end{figure}
\begin{figure}[!h]
\begin{center}
\includegraphics[width=15.0cm]{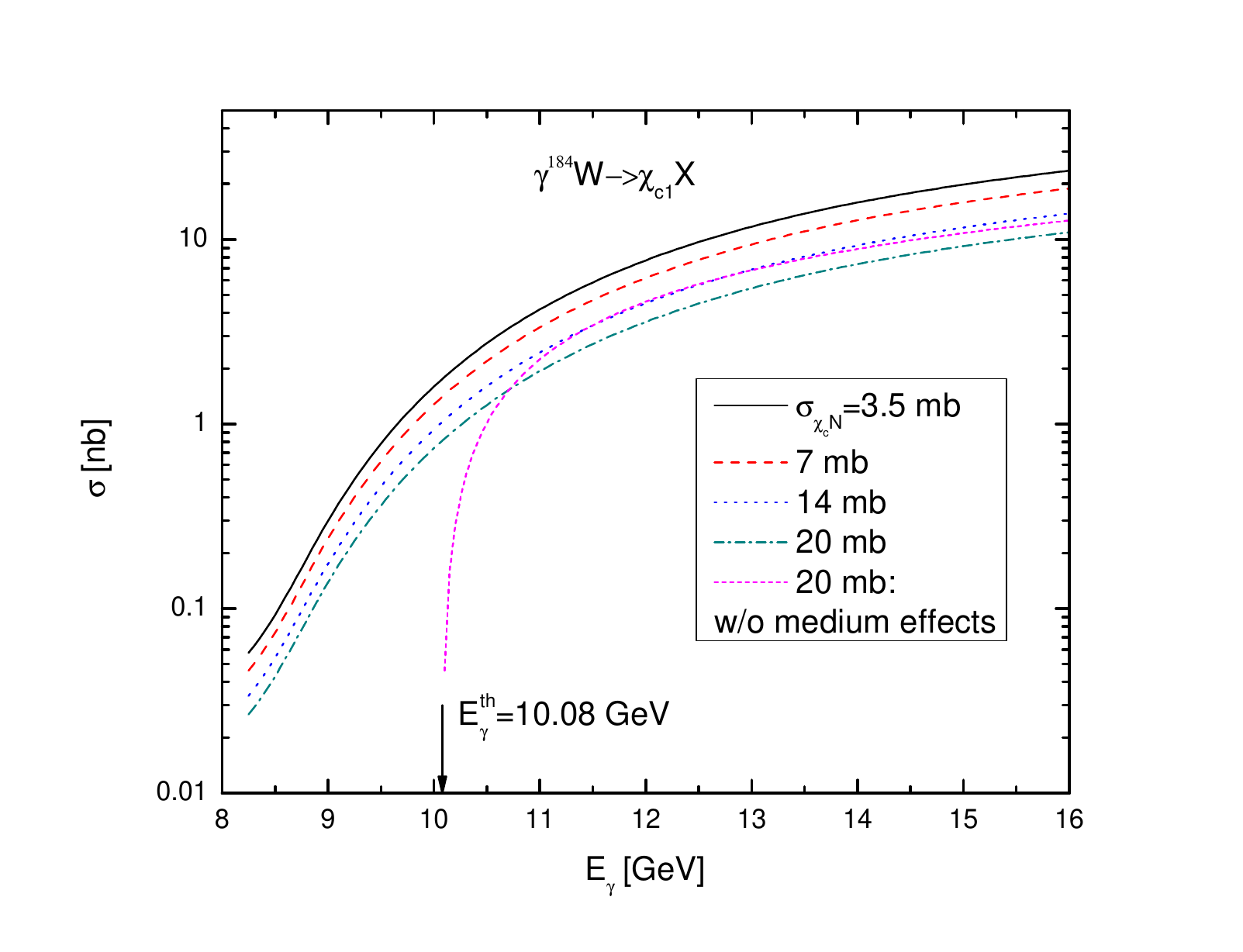}
\vspace*{-2mm} \caption{(Color online.) The same as in Fig. 3, but for the $^{184}$W target nucleus.}
\label{void}
\end{center}
\end{figure}

The information on the $\chi_{c}$ absorption cross section $\sigma_{\chi_{c}N}$ can also be extracted
from the comparison of the measured and calculated momentum distributions of $\chi_{c1}(1P)$ mesons from nuclei
in the photon energy range of interest.
Therefore, we consider now the momentum-dependent inclusive differential cross section for their production
with momentum $p_{\chi_{c1}}$ from the direct processes (1) and (2) in ${\gamma}A$ interactions.
Due to the fact that the $\chi_{c1}(1P)$ meson moves in the nucleus essentially forward in the lab frame
\footnote{$^)$Thus, for example, the maximum angle of its production on a free target proton at rest in
reaction (1) is about 6.4$^{\circ}$ at photon energy of 13 GeV.}$^)$,
we will calculate the $\chi_{c1}(1P)$ momentum distribution from considered target nuclei
for the laboratory solid angle ${\Delta}{\bf \Omega}_{\chi_{c1}}$ = $0^{\circ} \le \theta_{\chi_{c1}} \le 10^{\circ}$,
and $0 \le \varphi_{\chi_{c1}} \le 2{\pi}$. Then, according to the results presented both in Ref. [107] and above by
Eq. (3), we can get the following expression for this distribution:
\begin{equation}
\frac{d\sigma_{{\gamma}A\to {\chi_{c1}}X}^{({\rm dir})}
(p_{\gamma},p_{\chi_{c1}})}{dp_{\chi_{c1}}}=
2{\pi}I_{V}[A,\sigma_{{\chi_{c}}N}]
\int\limits_{\cos10^{\circ}}^{1}d\cos{{\theta_{\chi_{c1}}}}
\left<\frac{d\sigma_{{\gamma}p\to {\chi_{c1}}{p}}(p_{\gamma},
p_{\chi_{c1}},\theta_{\chi_{c1}})}{dp_{\chi_{c1}}d{\bf \Omega}_{\chi_{c1}}}\right>_A,
\end{equation}
where
$\left<\frac{d\sigma_{{\gamma}p \to {\chi_{c1}}p}(p_{\gamma},
p_{\chi_{c1}},\theta_{\chi_{c1}})}{dp_{\chi_{c1}}d{\bf \Omega}_{\chi_{c1}}}\right>_A$
is the off-shell differential cross section for production of $\chi_{c1}(1P)$ mesons
with momentum ${\bf p}_{\chi_{c1}}$ in the process (1),
averaged over the Fermi motion and binding energy of the intranuclear protons.
It can be expressed by Eqs. (28), (31)--(39) from Ref. [116], in which one needs to make the
substitution: $\Upsilon(1S) \to \chi_{c1}$. For brevity, we do not give
these expressions here. In order to calculate the c.m. $\chi_{c1}(1P)$ angular distribution in process (1)
(cf. Eq. (34) from Ref. [116]) one needs to know its exponential $t$-slope parameter $b_{\chi_{c1}}$ in
the threshold energy region. We adopt for this parameter the value of
2.0 GeV$^{-2}$, corresponding [107] to the $\psi(2S)$ slope parameter $b_{\psi(2S)}$ in the reaction
${\gamma}p \to \psi(2S)p$ at incident photon energy of 13 GeV.
\begin{figure}[!h]
\begin{center}
\includegraphics[width=15.0cm]{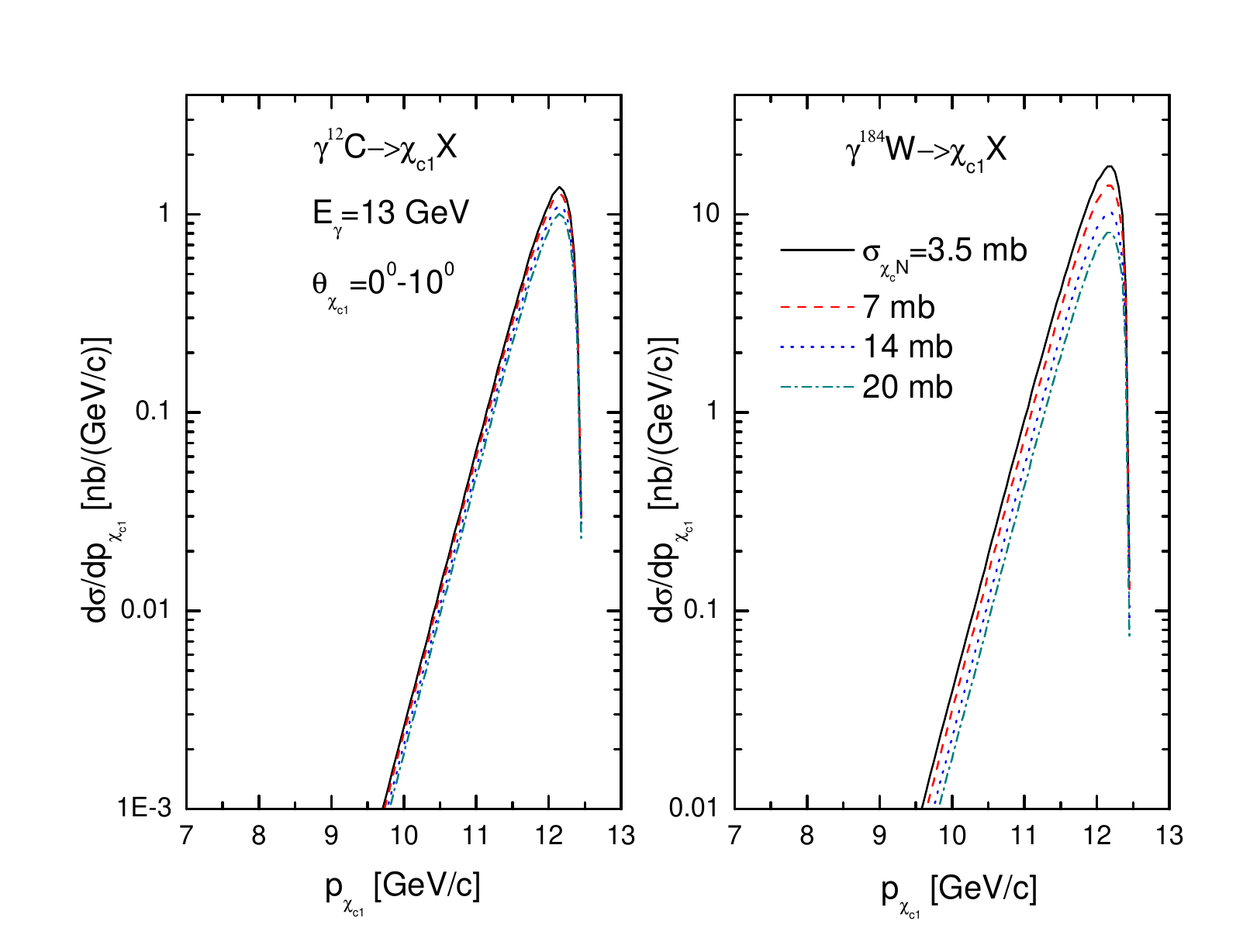}
\vspace*{-2mm} \caption{(Color online.) Momentum differential cross sections for the production of
$\chi_{c1}(1P)$ mesons from the direct processes (1), (2) in the laboratory polar angular range of
0$^{\circ}$--10$^{\circ}$ in the interaction of photons having energy of $E_{\gamma}=$ 13 GeV with $^{12}$C
(left) and $^{184}$W (right) nuclei, calculated for different values of the absorption cross section
$\sigma_{\chi_{c}N}$ indicated in the inset.}
\label{void}
\end{center}
\end{figure}
\begin{figure}[!h]
\begin{center}
\includegraphics[width=15.0cm]{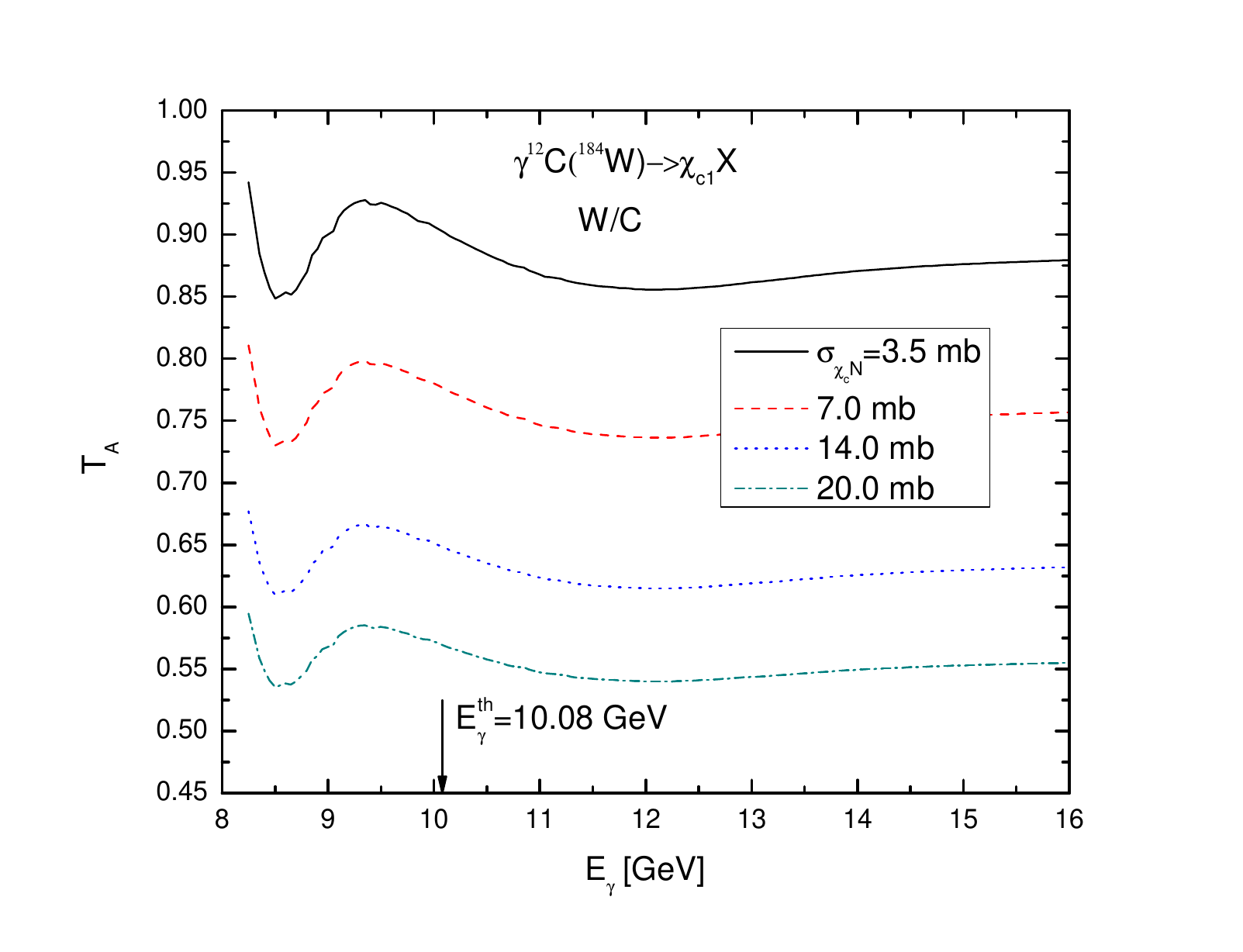}
\vspace*{-2mm} \caption{(Color online.) Transparency ratio $T_A$ for the $\chi_{c1}(1P)$ mesons
from the direct processes (1), (2) proceeding on an off-shell target nucleons
as a function of the incident photon energy for combination $^{184}$W/$^{12}$C, calculated for different
values of the absorption cross section $\sigma_{\chi_{c}N}$ indicated in the inset.
The arrow indicates the threshold energy for the $\chi_{c1}(1P)$ photoproduction on a free target
nucleon at rest.}
\label{void}
\end{center}
\end{figure}
\begin{figure}[!h]
\begin{center}
\includegraphics[width=15.0cm]{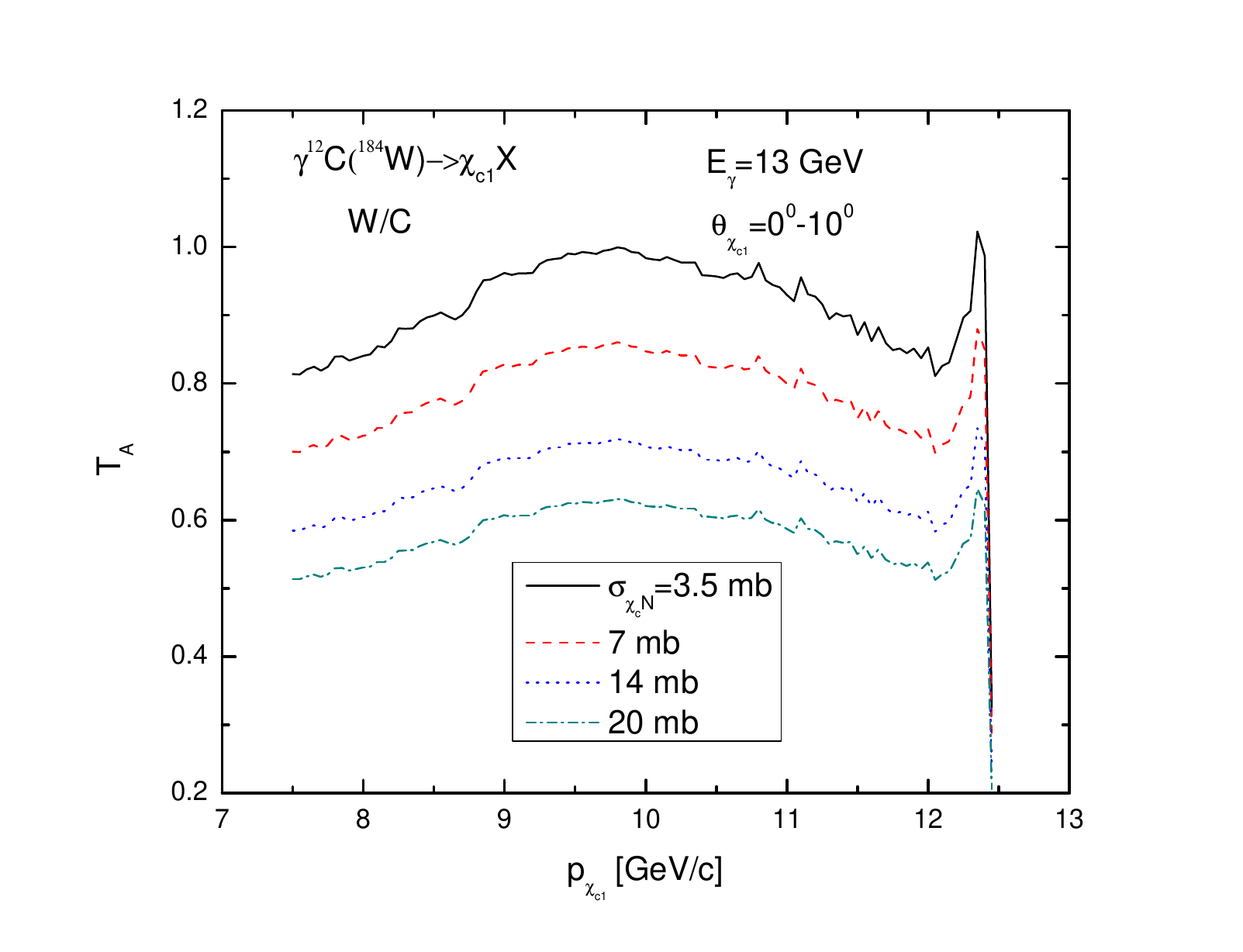}
\vspace*{-2mm} \caption{(Color online.) Transparency ratio $T_A$ for the $\chi_{c1}(1P)$ mesons
from the direct processes (1), (2) proceeding on an off-shell target nucleons
as a function of the $\chi_{c1}(1P)$ laboratory momentum for incident photon energy of 13 GeV for combination $^{184}$W/$^{12}$C, calculated in the laboratory polar angular range of 0$^{\circ}$--10$^{\circ}$
for different values of the absorption cross section $\sigma_{\chi_{c}N}$ indicated in the inset.}
\label{void}
\end{center}
\end{figure}
\begin{figure}[!h]
\begin{center}
\includegraphics[width=15.0cm]{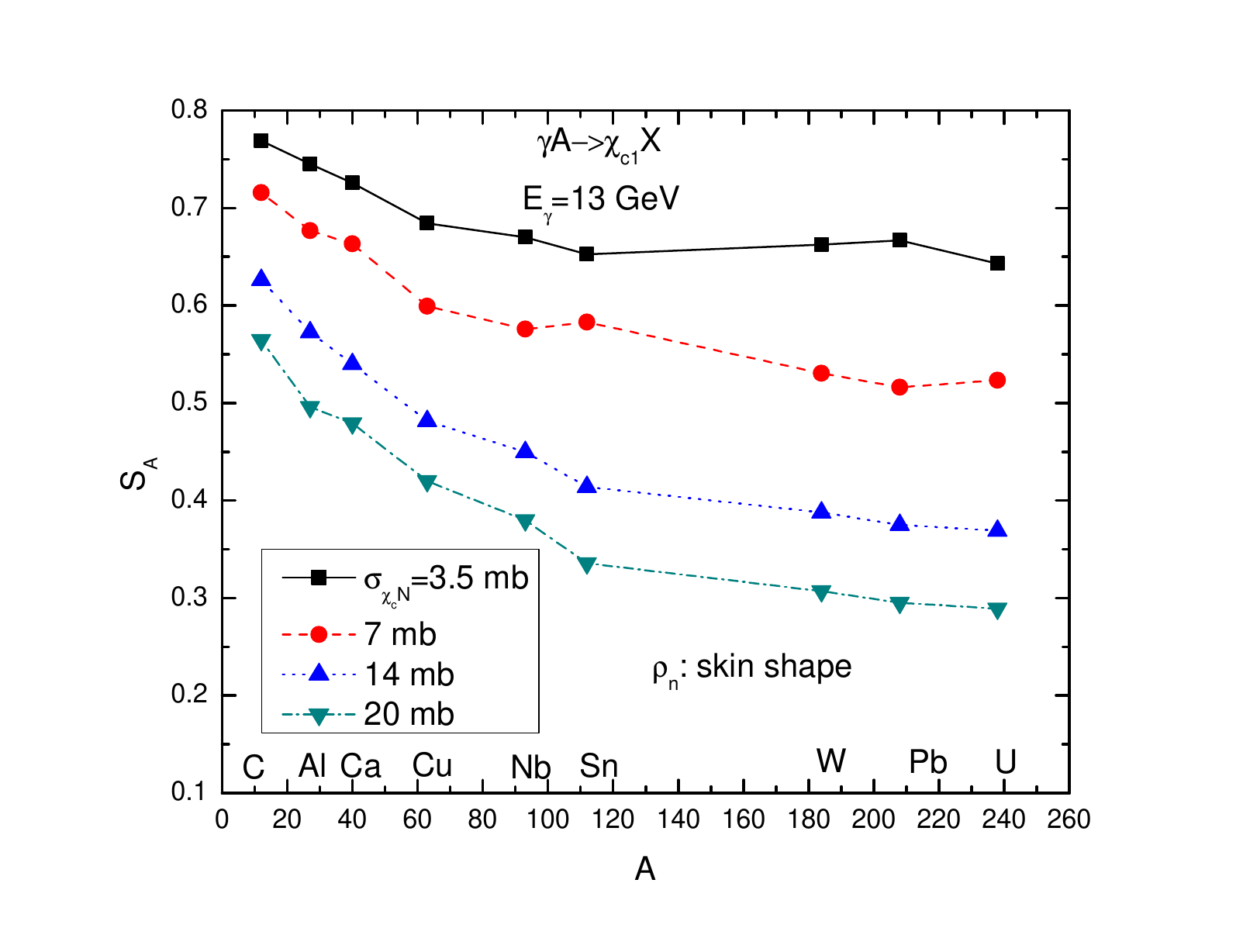}
\vspace*{-2mm} \caption{(Color online.) Transparency ratio $S_A$ for the $\chi_{c1}(1P)$ mesons
from the direct processes (1), (2) proceeding on an off-shell target nucleons
at incident photon energy of 13 GeV in the laboratory system as a function of the nuclear mass number $A$,
calculated for different values of the absorption cross section $\sigma_{\chi_{c}N}$ indicated in the inset.
The lines are to guide the eyes.}
\label{void}
\end{center}
\end{figure}
\begin{figure}[!h]
\begin{center}
\includegraphics[width=15.0cm]{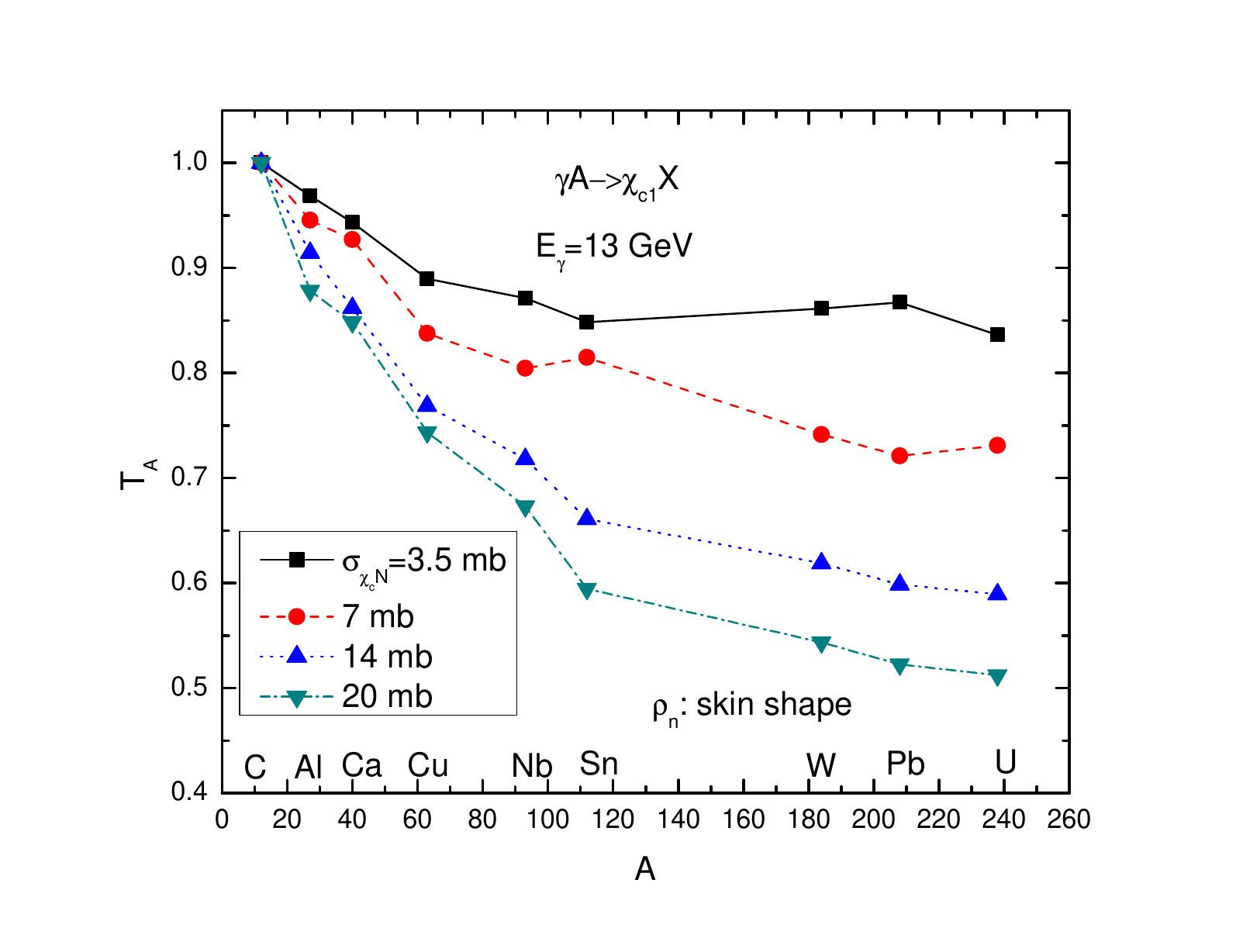}
\vspace*{-2mm} \caption{(Color online.) The same as in Fig. 8, but for the
transparency ratio $T_A$.}
\label{void}
\end{center}
\end{figure}
\begin{figure}[!h]
\begin{center}
\includegraphics[width=15.0cm]{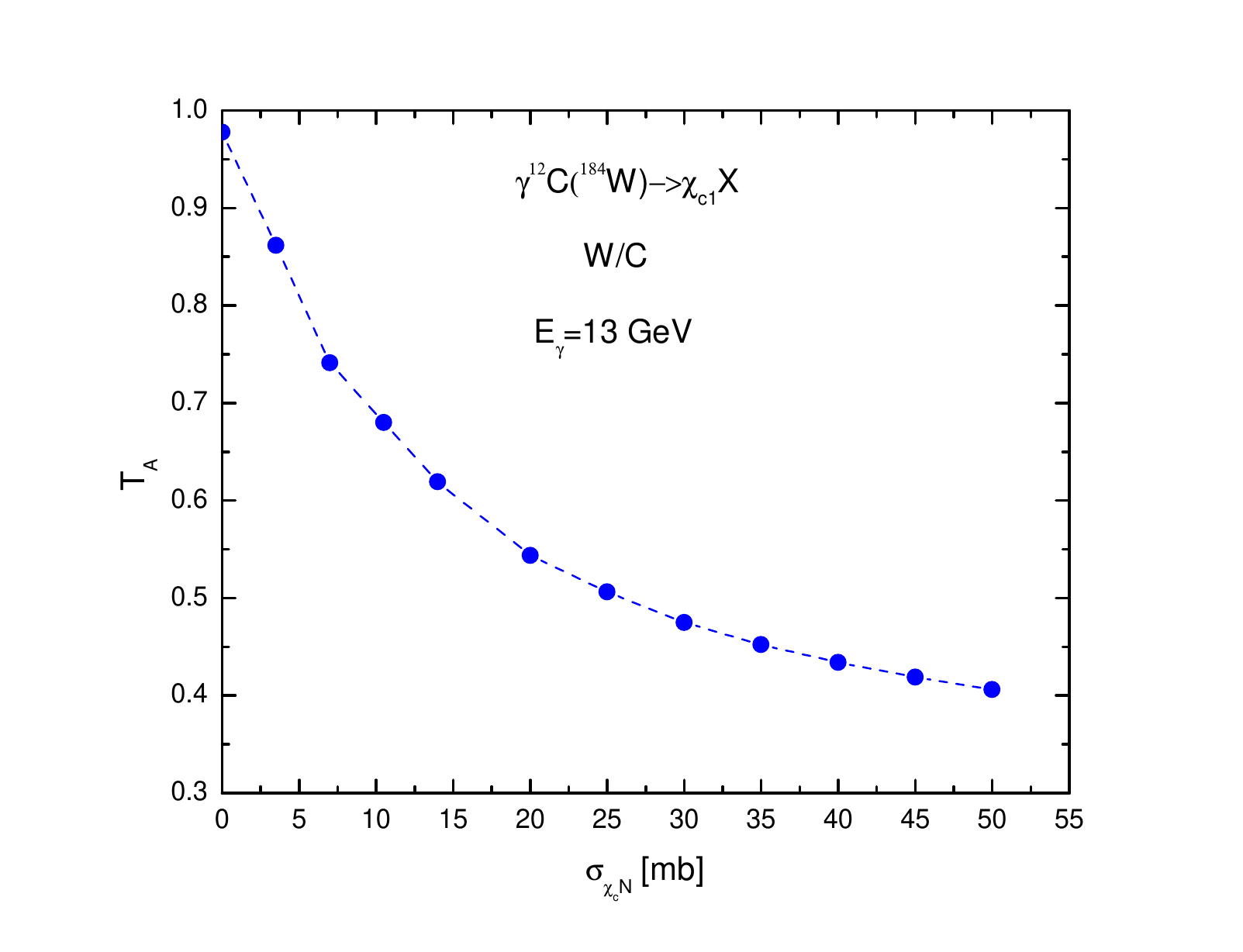}
\vspace*{-2mm} \caption{(Color online.) Transparency ratio $T_A$ for the $\chi_{c1}(1P)$ mesons
from the direct processes (1), (2) proceeding on an off-shell target nucleons
as a function of the absorption cross section $\sigma_{\chi_{c}N}$
at incident photon energy of 13 GeV for combination $^{184}$W/$^{12}$C.}
\label{void}
\end{center}
\end{figure}

\section*{3. Predictions and discussions}

\hspace{1.5cm} The excitation functions for production of $\chi_{c1}(1P)$ mesons on $^{12}$C and $^{184}$W nuclei,
calculated in line with Eq. (3) for four adopted options for the $\chi_c$ absorption cross section $\sigma_{\chi_{c}N}$ as well as for an off-shell target nucleons and for a free ones being at rest with the value of $\sigma_{{\chi_c}N}=20$ mb, are shown in Figs. 3 and 4, respectively. It is seen that the influence of the target nucleon Fermi motion on the $\chi_c1(1P)$ yield is significant at subthreshold photon energies ($E_{\gamma} < 10.08$ GeV).
It is also seen yet that for the heavy target nucleus $^{184}$W the obtained results depend more strongly on the $\chi_c$--nucleon absorption cross section than those for the light nucleus $^{12}$C. We observe for the $^{184}$W nucleus a well distinguishable and experimentally measurable differences $\sim$ 25--40\% between the results
corresponding to the calculations with employed values of the ${\chi_c}N$ absorption cross section.
For the light target nucleus $^{12}$C, the sensitivity of the cross sections to these values becomes lower and, respectively, the same differences as above become somewhat smaller. They are $\sim$ 10--15\% and will probably be
experimentally accessible as well in the future dedicated experiments at JLab, if the respective measurements will be performed with precision  better than 6--8\%. Such measurements look quite optimistic since the absolute values of
the $\chi_{c1}(1P)$ meson total photoproduction cross sections have a measurable strengths $\sim$ 0.3--2.0 nb and 3--20 nb for $^{12}$C and $^{184}$W target nuclei, respectively, at above threshold photon energies $\sim$ 11--16 GeV.
To motivate such measurements at 22 GeV CEBAF facility, it is desirable to estimate the $\chi_{c1}(1P)$ production rates (the event numbers) in the ${\gamma}^{12}$C and ${\gamma}^{184}$W reactions in a one-year run. For this aim, we translate the $\chi_{c1}(1P)$ photoproduction total cross sections, reported above, into the expected total cross sections of the
sequences ${\gamma}^{12}{\rm C}(^{184}{\rm W}) \to \chi_{c1}(1P)X$, $\chi_{c1}(1P) \to J/\psi{\gamma}$, $J/\psi \to e^+e^-$ by multiplying them on the appropriate branching ratios $Br[\chi_{c1}(1P) \to J/\psi{\gamma}]\approx$ 34.3\%,
$Br[J/\psi \to {e^+}{e^-}]\approx$ 6\% [81]. Then, to estimate the total numbers of the $\chi_{c1}(1P)$ events in a one-year run at the CEBAF facility, one needs to multiply the latter $\chi_{c1}(1P)$ "photoproduction" total cross sections on the carbon and tungsten target nuclei by the integrated luminosity of $\sim$ 500 pb$^{-1}$ [83]
as well as by the detection efficiency. Using a conservative detection efficiency of 10\%,
we estimate about of 310--2060 and 3100--20600 event numbers per year for the $\chi_{c1}(1P)$ signal in the cases of the $^{12}$C and $^{184}$W target nuclei, respectively. We see that a sufficiently large number of $\chi_{c1}(1P)$ events could be observed at well above threshold photon energies of 11--16 GeV. As a consequence, a future $\chi_{c1}(1P)$ excitation function measurements are, in principle, feasible and well motivated at these energies. They will allow for to set tight constraints on the $\sigma_{\chi_{c}N}$ cross section, once the $\chi_{c1}(1P)$ near-threshold proton (and neutron) target photoproduction cross sections will be experimentally known.

The momentum dependences of the absolute $\chi_{c1}(1P)$ meson differential cross sections from direct productions processes (1), (2) in the ${\gamma}^{12}$C and ${\gamma}^{184}$W collisions, calculated on the basis of Eq. (18) for four adopted values of the $\chi_c$--nucleon absorption cross section for laboratory angles of 0$^{\circ}$--10$^{\circ}$
and for initial photon energy of 13 GeV, are shown in Fig. 5. It can be clearly seen from this figure that the differential cross sections reveal a certain sensitivity
\footnote{$^)$Which is similar to that shown in Fig. 4.}$^)$
to this cross section mostly for the $^{184}$W target nucleus. Furthermore, in this case
the differential cross sections are roughly one order of magnitude larger than those on $^{12}$C nucleus and
they reach a rather measurable at the upgraded up to 22 GeV CEBAF facility strength $\sim$ 4--20 nb/(GeV/c) in the central momentum region of 11.5--12 GeV/c. Therefore, the $\chi_{c1}(1P)$ meson differential cross section measurements on heavy target nuclei at near-threshold photon energies $\sim$ 13 GeV will open yet another additional possibility to limit the ${\chi_c}N$ absorption cross section in cold nuclear matter.

To provide further guidance for future experiments we show in Figs. 6 and 7 the photon energy and the $\chi_{c1}$ meson momentum dependences of the transparency ratio $T_A$ for the $\chi_{c1}(1P)$ mesons for the $^{184}$W/$^{12}$C combination calculated in line with Eq. (17), using the results presented, respectively,
in Figs. 3, 4 and 5
\footnote{$^)$It should be noted that the definition of the transparency ratios $S_A$ and $T_A$ via equations (16) and (17) implies that they should be considered only at above threshold photon energies. But since the right-hand sides of these equations are defined both above and below the respective thresholds, we used them in calculating the transparency ratio $T_A$ also at subthreshold photon energies.}$^)$.
One can see that both dependences depend noticeably on the considered variations in the cross section $\sigma_{{\chi_c}N}$ as well. Thus, there are the measurable changes in them $\sim$ 15--20\% for these variations. Despite the fact that they are somewhat smaller than those between the calculations of the $\chi_{c1}$ total production cross section on $^{184}$W target nucleus (cf. Fig. 4), both considered dependences can also be used for discriminating between possible choices for the $\sigma_{{\chi_c}N}$ absorption cross section. Furthermore, they depend weakly on the photon energy and on the $\chi_{c1}$ momentum practically at all energies and momenta except of those belonging to the low-energy and to the high-momentum regions. These features can be used as well to better constrain the above absorption cross section.

The transparency ratios $S_A$ and $T_A$ of the $\chi_{c1}(1P)$ production from the direct processes (1), (2)
in ${\gamma}A$ reactions
($A=$$^{12}$C, $^{27}$Al, $^{40}$Ca, $^{63}$Cu, $^{93}$Nb, $^{112}$Sn, $^{184}$W, $^{208}$Pb, and $^{238}$U)
are depicted in Figs. 8 and 9, respectively, as functions of the mass number of the target nucleus.
They have been calculated for the photon beam energy of 13 GeV in line with Eqs. (16) and (17), correspondingly,
and for four adopted values of the $\chi_c$--nucleon absorption cross section $\sigma_{{\chi_c}N}$.
It is seen from these figures that the transparency ratios $S_A$ and $T_A$ depend strongly on variations in
this cross section and in the nuclear mass number. If we turn on absorption of $\chi_{c1}(1P)$ mesons, then they drop
strongly along the target nuclei $^{12}$C -- $^{238}$U and, in particular, the transparency ratios $T_A$ reach values
of the order of 0.5 for the heavy nuclei like $^{208}$Pb and $^{238}$U at $\sigma_{{\chi_c}N}=20$ mb -- a large deviation from unity which should be easily seen in a future experiment.
On the other hand, there are a sizeable and measurable variations $\sim$ 10, 23, 13\% in the ratio $S_A$
between calculations corresponding to the absorption cross sections of 3.5 and 7 mb, 7 and 14 mb, 14 and 20 mb, respectively,  already for relatively "light" nuclei like $^{40}$Ca.
For the medium-mass ($^{112}$Sn) and heavy ($^{238}$U) target nuclei these variations are even larger. They are
about 12, 40, 23\% and 23, 42, 28\%, respectively. So, the highest sensitivity of the quantity $S_A$ to the cross section $\sigma_{{\chi_c}N}$ is observed for heavy target nuclei.
For the quantity $T_A$ the analogous variations are smaller but yet are experimentally distinguishable in the range of large A. They are about 2, 8, 2\%, 4, 23, 10\% and 14, 24, 15\%, respectively, in the cases of relatively "light", medium-mass and heavy target nuclei mentioned above.
Hence, we can conclude that the observation of the A-dependences of the transparency ratios $S_A$ and $T_A$, at least, for large mass numbers $A$ in the future photoproduction experiments would definitely allow for the
discrimination between considered values of the ${\chi_c}N$ absorption cross section.

Another source of information about this cross section is shown in Fig. 10 the dependence of the $\chi_{c1}(1P)$ transparency ratio $T_A$ for the $^{184}$W/$^{12}$C combination on the size of the ${\chi_c}N$ absorption cross section calculated at photon energy of 13 GeV in line with Eq. (17) using the results presented in Figs. 3, 4 and additional ones obtained at this energy employing the additional options: 0, 10.5, 25, 30, 35, 40, 45 and 50 mb for this cross section. We see that it drops quickly with increasing this cross section up to 25 mb and reaches value of the order of 0.4 for its value of 50 mb -- a large deviation from unity which also should be easily seen in a future experiments and which give confidence that a comparison with experimental data will yield reliable information on the ${\chi_c}N$ absorption cross section.

In conclusion, our study identifies the absolute total and differential cross sections for near-threshold photonuclear production of unpolarized $\chi_{c1}(1P)$ mesons as well as their relative cross sections (transparency ratios) as a powerful observables for exploring the $\chi_{cJ}(1P)$ family ($J=0,1,2$) absorption in cold nuclear matter.
It provides valuable assistance and guidance for future experimental studies in this field [82, 117] at the upgraded up to 22 GeV CEBAF facility.

\section*{4. Conclusions}

\hspace{1.5cm} In view of the expected data on the photonuclear production of the 1$P$-wave heavy charmonium $\chi_{c1}(1P)$ from the JLab upgraded to 22 GeV, in the present work we have studied its inclusive photoproduction from nuclei near the kinematic threshold in the framework of the collision model, based on the nuclear spectral function, for incoherent direct photon--nucleon charmonium creation processes. The model accounts for the final $\chi_{c1}(1P)$ absorption in nuclear medium, target nucleon binding and Fermi motion. We have calculated the absolute and relative excitation functions on $^{12}$C and $^{184}$W target nuclei at near-threshold photon beam energies of 8.25--16.0 GeV, the absolute momentum differential cross sections and ratios of them for its production off these target nuclei at laboratory polar angles of 0$^{\circ}$--10$^{\circ}$ and for photon energy of 13 GeV as well as the A-dependences of the transparency ratios for the $\chi_{c1}(1P)$ at photon energy of 13 GeV within the different scenarios for its absorption cross section in nuclei. We demonstrate that the absolute and relative observables considered reveal a definite sensitivity to these scenarios. Therefore, they might be useful for the determination of this cross section
at finite momenta from the comparison of them with the experimental data from the future experiments at the upgraded up to 22 GeV CEBAF facility, providing valuable insights for theoretical and experimental studies of charmonium production and suppression in relativistic heavy-ion collisions.
\\

\end{document}